\documentclass[aps,showpacs,twocolumn,prb,longbibliography,nobibnotes]{revtex4-1}

\usepackage[dvipdfmx]{graphicx}
\usepackage{dcolumn}
\usepackage{bm}
\usepackage{hyperref}
\usepackage[utf8]{inputenc}
\usepackage{pslatex}
\usepackage{textcomp} 
\usepackage{color}
\usepackage{lmodern} 
\usepackage[normalem]{ulem}

\hypersetup{
        colorlinks = true,
        pdftitle = {}, 
        pdfauthor = {},
        pdfkeywords = {},
        linkcolor = blue,
        citecolor = blue,
        filecolor = black,
        urlcolor = magenta
}

\begin{document}

\title{Magnetic properties of Fe$_5$SiB$_2$ and its alloys with P, S, and Co}

\author{Miros\l{}aw Werwi\'nski}\email[Corresponding author: ]{werwinski@ifmpan.poznan.pl}
\affiliation{Department of Physics and Astronomy, Uppsala University, Box 516, SE-751 20 Uppsala, Sweden}
\affiliation{Institute of Molecular Physics Polish Academy of Sciences, ul. M. Smoluchowskiego 17, 60-179 Pozna\'{n}, Poland}
\author{Sofia Kontos}
\author{Klas Gunnarsson}
\author{Peter Svedlindh}
\affiliation{Department of Engineering Sciences, Uppsala University, Box 534, SE-751 21 Uppsala, Sweden}
\author{Johan Cedervall}
\author{Viktor H\"{o}glin}
\author{Martin Sahlberg}
\affiliation{Department of Chemistry, Uppsala University, Box 538, SE-751 21 Uppsala, Sweden}
\author{Alexander Edstr\"{o}m}
\author{Olle Eriksson}
\author{J\'{a}n Rusz}
\affiliation{Department of Physics and Astronomy, Uppsala University, Box 516, SE-751 20 Uppsala, Sweden}

\date{\today}

\begin{abstract}

Fe$_5$SiB$_2$ has been synthesized and magnetic measurements have been carried out, revealing that $M_{\text{sat}}$ = 0.92~MA/m at $T=300$~K. The $M$ vs $T$ curve shows a broad peak around $T=160$~K. The anisotropy constant, $K_1$, estimated at $T=300$~K, is 0.25~MJ/m$^3$. Theoretical analysis of Fe$_5$SiB$_2$ system has been carried out and extended to the full range of Fe$_5$Si$_{1-x}$P$_x$B$_2$, Fe$_5$P$_{1-x}$S$_x$B$_2$, and (Fe$_{1-x}$Co$_x$)$_5$SiB$_2$ compositions. The electronic band structures have been calculated using the Full-Potential Local-Orbital Minimum-Basis Scheme (FPLO-14). The calculated total magnetic moments are 9.20, 9.15, 9.59 and 2.42$\mu_B$ per formula units of Fe$_5$SiB$_2$, Fe$_5$PB$_2$, Fe$_5$SB$_2$, and Co$_5$SiB$_2$, respectively. 
In agreement with experiment, magnetocrystalline anisotropy energies (MAE's) calculated for T~=~0~K changes from a negative (easy-plane) anisotropy $-0.28$~MJ/m$^3$ for Fe$_5$SiB$_2$ to the positive (easy-axis) anisotropy 0.35~MJ/m$^3$ for Fe$_5$PB$_2$. 
Further increase of the number of p-electrons in Fe$_5$P$_{1-x}$S$_x$B$_2$ leads to an increase of MAE up to $0.77$~MJ/m$^3$ for the hypothetical Fe$_5$P$_{0.4}$S$_{0.6}$B$_2$ composition.
Volume variation and fixed spin moment calculations (FSM) performed for Fe$_5$SiB$_2$ show an inverse relation between MAE and magnetic moment in the region down to about 15\% reduction of the spin moment. The alloying of Fe$_5$SiB$_2$ with Co is proposed as a practical realization of magnetic moment reduction, which ought to increase MAE. MAE calculated in virtual crystal approximation (VCA) for a full range of (Fe$_{1-x}$Co$_x$)$_5$SiB$_2$ compositions reaches the maximum value of 1.16~MJ/m$^3$ at Co concentration $x$~=~0.3, with the magnetic moment 7.75$\mu_B$ per formula unit. Thus, (Fe$_{0.7}$Co$_{0.3}$)$_5$SiB$_2$ is suggested as a candidate for a rare-earth free permanent magnet. For the stoichiometric Co$_5$SiB$_2$ there is an easy-plane magnetization, with the value of MAE~=~$-0.15$~MJ/m$^3$.
\end{abstract}


\maketitle

\section{Introduction}

Volatile prices of the rare-earth elements observed in the recent years have motivated wide research initiatives in an effort to find replacement materials for the permanent magnets industry, which would use little or none of the rare-earth elements. Late $3d$ transition elements, such as Mn, Fe or Co are the primary candidates for such materials, thanks to their large and stable magnetic moments and high magnetic transition temperatures found in many of the compounds they form. 
However, weak spin-orbital interaction strength typically causes low magnetocrystalline anisotropy energy (MAE), which in turn means that they often lack appreciable coercivity, forming soft magnetic materials. 
Under certain circumstances, however, the MAE of 3$d$ transition metal compounds can be substantial, for example the tetragonally distorted Fe/Co alloys~\cite{Burkert04,Andersson06,Delczeg14,Reichel,Reichel15}, L1$_0$ phases such as FeNi, CoNi, MnAl~\cite{Miura13,Edstrom14}, Fe$_{16}$N$_2$~\cite{Kim72}, (Fe/Co)$_2$B~\cite{Iga70,Belashchenko15,Edstrom15} or Fe$_2$P-based alloys~\cite{Fujii77,Delczeg10}.
A general overview on rare-earth-free permanent magnets is available in Ref.~\cite{Niarchos15}.
All of the above mentioned systems are hexagonal or tetragonal, i.e., uniaxial.
For this reason, in this work we have focused on an iron-based system Fe$_5$SiB$_2$ and its alloys (Fe$_{1-x}$Co$_x$)$_5$SiB$_2$ and Fe$_5$Si$_{1-x}$P$_x$B$_2$ - they are tetragonal and rich on magnetic elements.

Fe$_5$SiB$_2$~\cite{Aronsson60} and Fe$_5$PB$_2$~\cite{Rundqvist62} are isomorphous with Cr$_5$B$_3$. They were initially reported by Aronsson and Engstr\"{o}m~\cite{Aronsson60} and Rundqvist~\cite{Rundqvist62} in 1960 and 1962. Several other borides crystallize within the same structure type. In the more general formula $X_5$SiB$_2$ for the silico-borides, $X$ may stand for (Mn, Fe, Mo)\cite{Aronsson60}, V~\cite{Reis07}, Nb~\cite{Brauner09} or W~\cite{Fukuma11}. The recently discovered Nb$_5$SiB$_2$ and W$_5$SiB$_2$ have been recognized as superconductors. Isomorphous phases with phosphorus, $X_5$PB$_2$, were identified with $X =$ (Mn, Fe, Co)~\cite{Rundqvist62}, Cr~\cite{Baurecht71} and Mo~\cite{Ilnitskaya85}.

\begin{figure}[ht]
\includegraphics[width=\columnwidth]{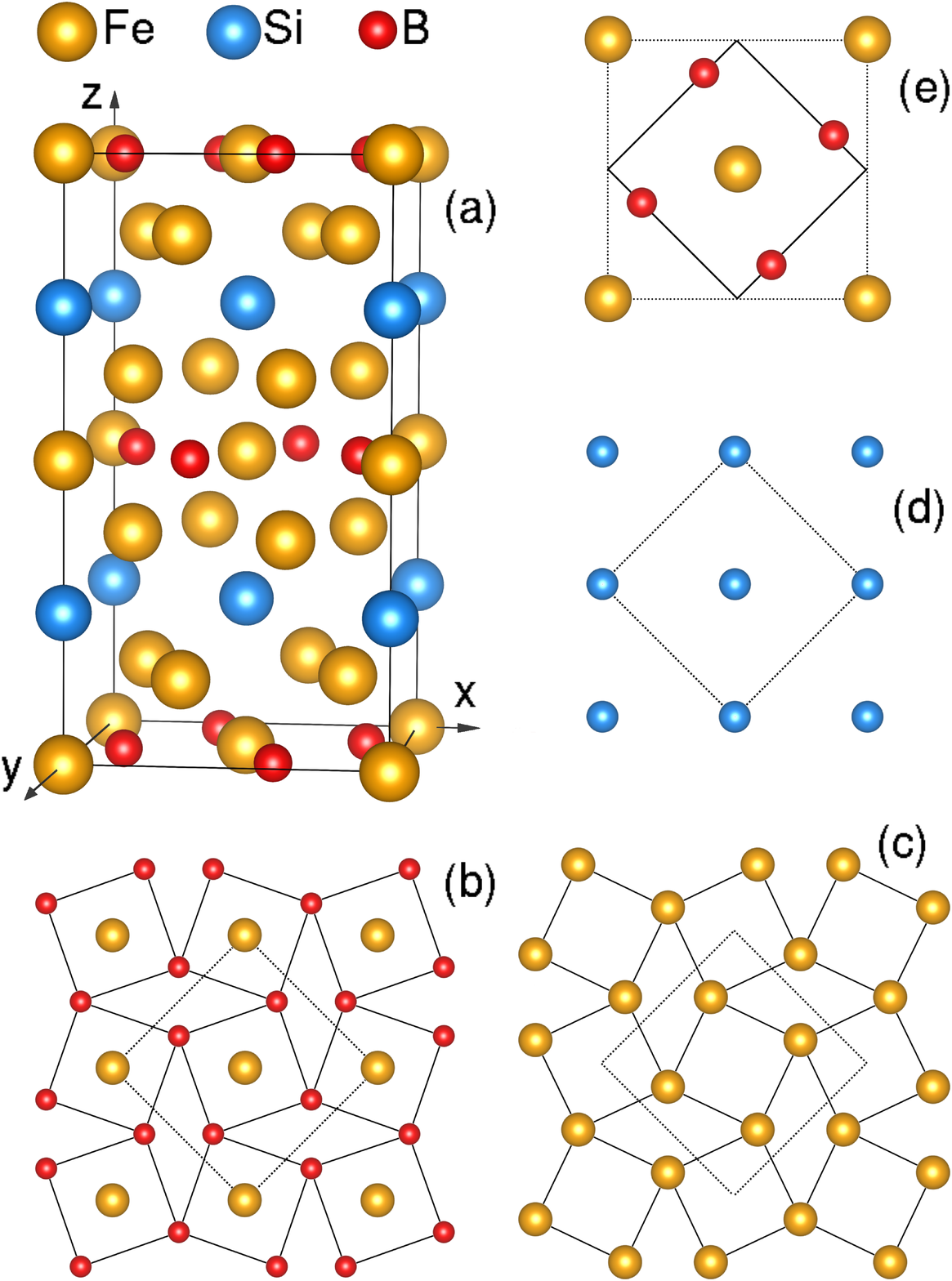}
\caption{\label{fig:structure} (a) The crystallographic structure of Fe$_5$SiB$_2$; (b) Fe$_2$-B layer; (c) Fe$_1$ layer; (d) Si layer; (e) Fe$_2$-B layer limited to the unit cell boundaries.}
\end{figure}

The crystal structure of Fe$_5$PB$_2$ was determined by Rundqvist~\cite{Rundqvist62} and Fe$_5$SiB$_2$ by Aronsson and Engstr\"{o}m~\cite{Aronsson60}, see Fig.~\ref{fig:structure}. Rundqvist stated that the distribution of P and B atoms in Fe$_5$PB$_2$ was ordered and had a narrow homogeneity range~\cite{Rundqvist62}. Both Fe$_5$PB$_2$ and Fe$_5$SiB$_2$ crystallize in the Cr$_5$B$_3$-type structure~\cite{Rundqvist62} with body-centered tetragonal (bct) unit cell, space group I4/mcm, see Fig.~\ref{fig:structure}(a). Single unit cell of Fe$_5X$B$_2$ contains 32 atoms divided into 4 formula units. Fe$_1$ atoms are distributed on the 16-fold position, B on the 8-fold, Fe$_2$ and Si/P on the 4-fold position, respectively.

The Fe$_5$SiB$_2$ and Fe$_5$PB$_2$ structures have a principally ordered arrangement of Si/P and B on their respective crystallographic sites, but some intermixing occurs as well as interstitial accommodations of B, Si and P. Furthermore, octahedral holes of sufficient size for small non-metal atoms exist in the system. Aronsson and Engstr\"{o}m~\cite{Aronsson60} found it likely that some of the boron atoms may be interstitially accommodated in these holes. Moreover, Rundqvist~\cite{Rundqvist62} described two types of holes in the Cr$_5$B$_3$-type lattice, trigonal prismatic and anti-prismatic ones. In the Fe$_5$PB$_2$ system the radii of the trigonal prismatic and anti-prismatic holes are 0.93~\AA{} and 1.08~\AA{}, respectively, while the B radius is 0.88~\AA{}. A homogeneity range of Fe$_5$SiB$_2$ with the substitutions of B for Si has been observed previously. Fe$_5$SiB$_2$ measured by Aronsson and Engstr\"{o}m~\cite{Aronsson60} have B/Si ratio slightly larger than 2.00. The authors even proposed a new 
general formula, Fe$_5$Si$_{1-x}$B$_{2+x}$ ($0<x<0.1$).
Another feature of Fe$_5X$B$_2$ structure is its layered character. The structure is built up of three types of layers, which alternate along the $c$-axis. A single unit cell consists of 8 layers: 2 Fe-B, 4 Fe, and 2 Si layers, see Fig.~\ref{fig:structure}.

At room temperature Fe$_5$SiB$_2$ and Fe$_5$PB$_2$ are ferromagnets with an orientation of spins parallel to the tetragonal axis~\cite{Ericsson78}. The Curie temperature ($T_C$) is slightly varying with the deviation from the stoichiometry (Fe$_5X_{1-x}$B$_{2+x}$), $T_C$~=~628$\pm$11~K for Fe$_5$PB$_2$~\cite{Haggstrom75} and $T_C$~$\sim$~784~K for Fe$_5$SiB$_2$~\cite{Ericsson78}.
For Fe$_5$PB$_2$ the tetragonal easy axis of the magnetization direction is valid from 80~K up to $T_C$~\cite{Ericsson78}. The low temperature M\"{o}ssbauer study of Fe$_5$SiB$_2$ revealed a significant change of the spectral shape in the temperature range of 130 to 150~K~\cite{Ericsson78}. Ericsson et al.~\cite{Ericsson78} concluded that below 140~K the spins were oriented in the $ab$-plane or close to that. There were also indications that below 140~K the pure ferromagnetic coupling is partly destroyed~\cite{Ericsson78}. In the temperature region from 140~K up to $T_C$ the Fe$_5$SiB$_2$ was confirmed to be a ferromagnet with the spins parallel to the $c$-axis~\cite{Ericsson78}.
H\"{a}ggstr\"{o}m et al.~\cite{Haggstrom75} referred that an average magnetic moment of Fe$_5$PB$_2$ was reported to be $1.73\mu_B$~per~Fe~atom. The individual moments calculated by H\"{a}ggstr\"{o}m et al. from the values of the magnetic hyperfine fields are $1.6\mu_B$ for Fe$_1$ and $2.2\mu_B$ for Fe$_2$.

To the best of our knowledge there is no literature data concerning Co$_5$SiB$_2$, but only its homologue Co$_5$PB$_2$~\cite{Rundqvist62}. Very recently, Co- and Mn-doped Fe$_5$SiB$_2$ and Fe$_5$PB$_2$ were synthesized and characterized\cite{McGuire15}. Another recent manuscript reports a study of single crystal of Fe$_5$PB$_2$\cite{Lamichhane16}. Both these studies also report basic characterization of the electronic structure of the parent compounds, Fe$_5$SiB$_2$ and Fe$_5$PB$_2$, respectively.

\section{Experimental and computational details}\label{sec:exp_comp_details}

\subsection{Experimental details}\label{subsec:exp_details}

Fe$_5$SiB$_2$ samples were prepared with a conventional arc furnace in an argon atmosphere. Stoichiometric amounts of iron (Leico Industries, purity 99.995~\%, surface oxides were reduced in H$_2$-gas), silicone (Highways International, purity 99.999~\%) and zone melted boron (Wacher) were used as raw materials. The samples were crushed, pressed into pellets and sealed in silica tubes, annealed at 1273~K for 19 days and then quenched in cold water.

X-ray powder diffraction intensities (XRD) were recorded using Bruker D8 diffractometer with a V\aa{}ntec position sensitive detector with a 4\textdegree{} opening using CuK$_{\alpha1}$ radiation, $\lambda = 1.540598$~\AA{}. The measurements were taken in a 2$\theta$-range of 20-90\textdegree{}. Phase analysis and crystal structures refinements were performed using the Rietveld method~\cite{Rietveld69} implemented in the software FullProf~\cite{Rodriguez93} and cell parameters were refined using the software UNITCELL~\cite{Holland97}. 

Magnetization measurements on the powder samples were performed using a Quantum Design PPMS 6000 Vibrating Sample Magnetometer. The low field magnetization temperature dependence was recorded between $T=300$~K and $T=10$~K at a constant magnetic field of 8~kA/m. The magnetization, $M$, vs field, $H$, at constant $T=300$~K was measured in the range from $-5.6$~MA/m to $+5.6$~MA/m.

\subsection{Computational details}\label{subsec:comp_details}

Fully relativistic electronic band structure calculations were carried out by using the Full-Potential Local-Orbital Minimum-Basis Scheme (FPLO-14)~\cite{Koepernik99} within the generalized gradient approximation (GGA) for the exchange-correlation potential in the Perdew, Burke, Ernzerhof form (PBE)~\cite{Perdew96}. Calculations were performed with $16 \times 16 \times 16$ \textbf{k}-mesh and convergence criterion $10^{-8}$~Ha. An initial spin splitting is applied to the $3d$-atoms, assuming ferromagnetic structures, as observed experimentally.

\begin{table}[!ht]
\caption{\label{tab:crystal_data} The calculated optimized crystallographic parameters for Fe$_5$SiB$_2$, Co$_5$SiB$_2$ and Fe$_5$PB$_2$. Space group I4/mcm, no. 140.}
\centering
\begin{tabular}{l|ccccc}
\hline \hline
system        & $a$ [\AA{}] & $c$ [\AA{}] & $x_{\text{Fe$_1$/Co$_1$}}$ & $z_{\text{Fe$_1$/Co$_1$}}$ & $x_{\text{B}}$ \\
\hline
Fe$_5$SiB$_2$ &   5.546     &   10.341    &   0.169          &   0.136          & 0.382 \\
Co$_5$SiB$_2$ &   5.511     &    9.953    &   0.169          &   0.134          & 0.376 \\
Fe$_5$PB$_2$  &   5.503     &   10.347    &   0.172          &   0.138          & 0.383 \\
Fe$_5$SB$_2$  &   5.466     &   10.594    &   0.172          &   0.138          & 0.383 \\
\hline \hline
\end{tabular}
\end{table}

For Fe$_5$SiB$_2$, Co$_5$SiB$_2$, Fe$_5$PB$_2$, and Fe$_5$SB$_2$ the lattice parameters and the internal atomic positions were optimized, see Table~\ref{tab:crystal_data}. The optimized structural parameters of Fe$_5$SiB$_2$ and Fe$_5$PB$_2$ are in an excellent agreement with the corresponding experimental values~\cite{Ericsson78, Rundqvist62}. We note that the computationally relaxed structural parameters reported here are for the stoichiometric phases of Fe$_5$SiB$_2$ and Fe$_5$PB$_2$, while the previously measured values are estimated from non-fully stoichiometric samples.

Virtual crystal approximation (VCA) is used to study three ranges of compositions, the first one between Fe$_5$SiB$_2$ and Co$_5$SiB$_2$, the second one between Fe$_5$SiB$_2$ and Fe$_5$PB$_2$, and the last one between Fe$_5$PB$_2$ and Fe$_5$SB$_2$. The general idea of the VCA is to imitate a homogeneous random on-site occupation of two types of atoms by using a virtual atom with a corresponding averaged value of the nuclear charge and number of electrons.
Application of VCA for Fe/Co alloying is motivated by the fact that Co atomic number is one above Fe and furthermore that similar VCA calculations have already produced valuable results for other Fe/Co alloys~\cite{Delczeg14, Edstrom15, Reichel15}.
Since P and Si are also neighbors in the periodic table, and as the $p$-elements are basically supplying itinerant electrons to the crystal, the VCA is expected to be a good approximation of the substitutional disorder also for this class of systems. The same argument holds for P/S alloying.
It is important to comment that although VCA is expected to give the right qualitative trends of MAE changes with Fe/Co concentration, the exact values of MAE are usually overestimated by a factor between 2 and 4 in comparison to experiment. 
More accurate results of MAE can be calculated within the coherent potential approximation (CPA)~\cite{Soven70,Turek12,Kota12,Edstrom15}.
Unfortunately, the utilized version of the FPLO code, which allows to perform the spin-orbit coupling (SOC) calculations together with fixed spin moment (FSM), does not allow to combine SOC with CPA.
Calculations for Fe$_5$Si$_{1-x}$P$_x$B$_2$ compositions start from the Fe$_5$SiB$_2$ structure with a full Si site occupation. The Si atom is then replaced by a virtual atom to mimic the specific composition. The atomic number of virtual atom is $Z_{\text{VCA}} = (1-x) \cdot Z_{\text{Si}} + x \cdot Z_{\text{P}}$. The lattice parameters of the intermediate compositions are interpolated linearly between the lattice parameters of the extreme systems, $a_{\text{VCA}} = (1-x) \cdot a_{\mathrm{Fe}_5\mathrm{SiB}_2} + x \cdot a_{\mathrm{Fe}_5\mathrm{PB}_2}$. For every composition the internal atomic positions are optimized.
The VCA calculations for (Fe$_{1-x}$Co$_x$)$_5$SiB$_2$ series are performed with lattice parameters and internal atomic positions interpolated as the linear combinations of crystal parameters of boundary compositions Fe$_5$SiB$_2$ and Co$_5$SiB$_2$.
 
VESTA code~\cite{Momma08} was used for visualization of crystal structures.

\section{Experimental results}\label{sec:exp}

\subsection{Phase analysis and crystal structure}

\begin{figure}[ht]
\includegraphics[width=\columnwidth]{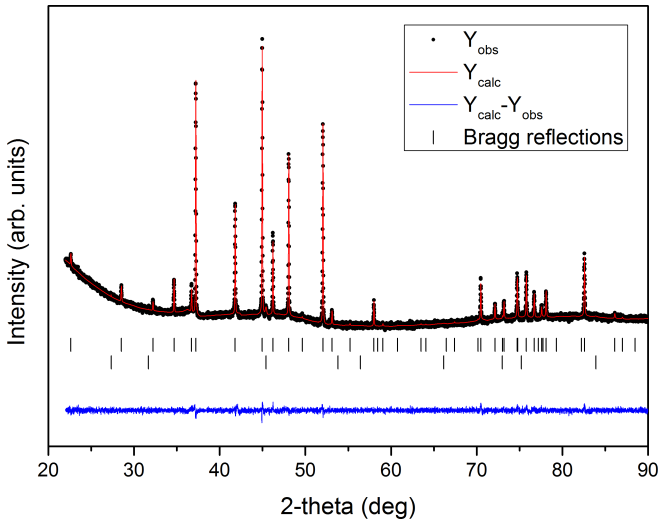}
\caption{\label{fig:xrd} 
X-ray powder diffraction pattern of Fe$_5$SiB$_2$ refined with the Rietveld method. The black dots and the red line correspond to the observed and calculated patterns, respectively, while the blue line shows the difference between observed and calculated data. Bars show the theoretical Bragg peaks of the Fe$_5$SiB$_2$ phase (upper) and the impurity phase, Fe$_3$Si (lower).}
\end{figure}

The XRD intensities, presented in Fig.~\ref{fig:xrd}, confirm that Fe$_5$SiB$_2$ crystallizes in the tetragonal Cr$_5$B$_3$-type structure with the space group I4/mcm~\cite{Aronsson60}. Synthesized sample contains less than 2~at.\% of an additional Fe$_3$Si impurity phase.

\begin{table}[ht]
\caption{\label{tab:wyckoff} 
Atomic coordinates and site occupancies for Fe$_5$SiB$_2$ derived from crystal structure refinement of XRD data.}
\begin{tabular}{l|l|ccc|l}
\hline \hline
Atom  	& Site	& $x$ 		& $y$  		 & $z$  	& occupancy\\
\hline
Fe$_1$  	& 16(l)	& 0.1683(1) 	& $\frac{1}{2}+x$& 0.1384(1) 	& 0.500\\
Fe$_2$   	& 4(c) 	& 0 		& 0 		 & 0 		& 0.125\\
Si  	& 4(a) 	& 0 		& 0		 & $\frac{1}{4}$& 0.125\\
B   	& 8(h) 	& 0.378(1) 	& $\frac{1}{2}+x$& 0 		& 0.250\\
\hline \hline
\end{tabular}
\end{table}

Table~\ref{tab:wyckoff} presents the atomic coordinates of Fe$_5$SiB$_2$ structure derived from crystal structure refinement of XRD data. Determined lattice parameters are $a=5.5561(3)$~\AA{} and $c=10.3476(10)$~\AA{}, estimated standard deviations are in the parenthesis. Previously reported Fe$_5$SiB$_2$ lattice parameters obtained by Ericsson~\cite{Ericsson78} are $a=5.43$~\AA{} and $c=10.33$~\AA{}, in close agreement with our results. The Fe$_5$PB$_2$ lattice parameters are $a=5.482$~\AA{} and $c=10.332$~\AA{}~\cite{Rundqvist62}. 
It indicates that there are only few promiles differences between the calculated and measured lattice parameters for Fe$_5$SiB$_2$ and Fe$_5$PB$_2$, compare with Table~\ref{tab:crystal_data}.

\subsection{Measured magnetic properties}

The value of $M_{\text{sat}}$ ($M$ at $H=5.6$~MA/m) at $T=300$~K is 0.92~MA/m ($6.3\mu_B$/f.u.) which is considerably smaller than expected from M\"{o}ssbauer measurements, 1.2~MA/m ($8.2\mu_B$/f.u.). However, the M\"{o}ssbauer data was reported for $T=80$~K. Further, due to surface effects, $M_{\text{sat}}$ for small particles should be lower than the bulk value. Also, there are non-magnetic impurities in the sample, e.g. Fe$_{3}$Si, decreasing the value of $M_{\text{sat}}$. Assuming the sample to consist of crystallites with uniaxial anisotropy, the anisotropy constant, $K_1$, was estimated using the $1/H^2$-term in the law of approach to saturation. This term is assumed to originate from the magnetization rotating against the magnetocrystalline anisotropy:
\begin{equation}
\label{eq:1}
\frac{M(H)}{M_{\text{sat}}} = \left ( 1 - \frac{a_{2}}{H^{2}} \right ),
\end{equation}
where 
\begin{equation}
a_{2} = \frac{4}{15} \left ( \frac{K_{1}}{M_{\text{sat}}} \right )^2
\end{equation}

Using the $M$ vs $H$ measurement at $T=300$~K, the result for Fe$_5$SiB$_2$ is $K_1=0.25$~MJ/m$^3$. Since Eq.~\ref{eq:1} contains only the $1/H^2$ dependence of $M(H)$, the selection of field interval for the K$_1$ estimation is crucial. Here, the field interval 0.6--1.0~MA/m was used, corresponding to $M(H)/M_{\text{sat}}= 0.95$--$0.98$. However, the need of a more accurate method, e.g. involving a single crystal sample, is emphasized and will be the aim for further experiments.

\begin{figure}[ht]
\includegraphics[trim = 80 0 0 60,clip,height=0.9\columnwidth,angle=270]{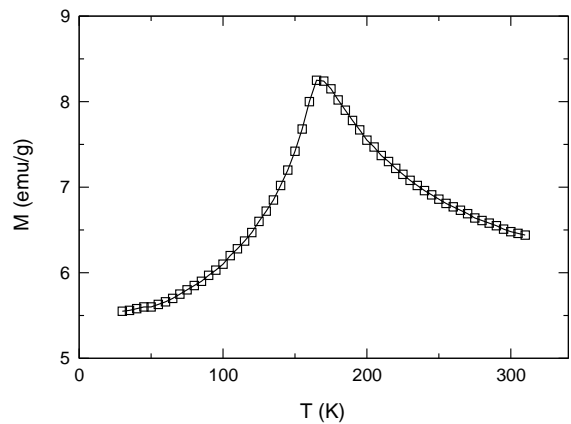}
\caption{\label{fig:fe5sib2_M_vs_T} The temperature dependency of magnetization as measured for Fe$_5$SiB$_2$.}
\end{figure}

Concerning the proposed reorientation of spins at $T=140$~K reported from M\"{o}ssbauer experiments, the following observation was made from magnetization measurements, see Fig.~\ref{fig:fe5sib2_M_vs_T}:
The $M$ vs $T$ curve shows a broad peak around $T=160$~K and then $M$ decreases continuously with decreasing temperature to an approximately constant level at $T=10$~K. The decrease in $M$ from $T=160$~K to $T=10$~K is 30\%.

\section{Density Functional Theory Calculations}\label{sec:dft}

\subsection{Fe$_5$SiB$_2$ Electronic Structure}

\begin{figure}[ht]
\includegraphics[trim = 0 5 85 90 ,clip,width=0.8\columnwidth]{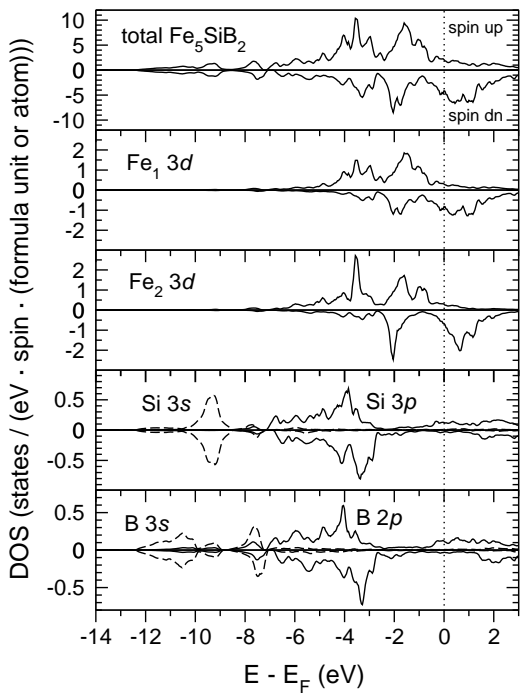}
\caption{\label{fig:dos} The spin projected partial and total densities of states DOS for Fe$_5$SiB$_2$. Calculations done within GGA(PBE)+so approximation.}
\end{figure}

The basic electronic configurations of the elements forming Fe$_5$SiB$_2$ are Fe: [Ar] $4s^2 3d^6$, Si: [Ne] $3s^2 3p^2$ and B: [He] $2s^2 2p^1$. This is also reflected in the densities of electronic states (DOS), see Fig.~\ref{fig:dos}. The total DOS is a sum of the individual atomic contributions, from which the dominant role is played by Fe $3d$ bands. Fe $3d$, Si $3s$, Si $3p$, B $2s$, and B $2p$ bands are presented on the spin projected DOS plots. The Fe bands Fe$_1$ $3d$ and Fe$_2$ $3d$, for two inequivalent Fe sites, are both strongly spin polarized. The corresponding spin magnetic moments are 1.87 and 2.24$\mu_B$/atom, respectively. The spin, orbital and total magnetic moments are summarized in Table~\ref{tab:mm}. The Si and B spin magnetic moments are both equal to $-0.25\mu_B$.

From $-2.5$~eV up to the Fermi level the character of DOS originates almost entirely from Fe$_1$ $3d$ and Fe$_2$ $3d$ bands. It means that the main contribution to the DOS at the Fermi level is provided by the Fe $3d$ states, what moreover leads to high spin polarization on the Fermi level, of about 66~\%.
Furthermore, calculations show that the orbital magnetic moments are present only on the Fe atoms (see Table~\ref{tab:mm}). Their values $0.039\mu_B$ and $0.048\mu_B$ are close to the value $0.043\mu_B$ for bcc iron, calculated within the same frame. Furthermore, the all theoretical values are reduced in comparison to the experimental value $0.086\mu_B$ for the bcc iron~\cite{Chen95}. The total magnetic moment per formula unit is equal to $9.20\mu_B$, which is in good agreement with the recently determined experimental value $9.35\mu_B$/f.u. ($5*1.87\mu_B$/Fe) from $M(H)$ measurements at 10~K~\cite{Cedervall16}.

\begin{table}[ht]
\caption{\label{tab:mm} The spin, orbital and total magnetic moments [$\mu_B$/(atom or formula unit)] for Fe$_5$SiB$_2$, Co$_5$SiB$_2$, Fe$_5$PB$_2$, and Fe$_5$SB$_2$.
}
\begin{tabular}{c|cc|cc|cc|cc}
\hline \hline
       &\multicolumn{2}{c|}{Fe$_5$SiB$_2$}&\multicolumn{2}{c|}{Co$_5$SiB$_2$}&\multicolumn{2}{c|}{Fe$_5$PB$_2$}&\multicolumn{2}{c}{Fe$_5$SB$_2$}\\
\hline       
atom   		& $\mu_s$ & $\mu_l$   & $\mu_s$ & $\mu_l$   & $\mu_s$ & $\mu_l$  & $\mu_s$ & $\mu_l$ \\
\hline
Fe$_1$/Co$_1$ 	&  1.87   &   0.039   &  0.40   &  0.016    &  1.83   &  0.034   &  1.91   &  0.031 \\
Fe$_2$/Co$_2$ 	&  2.24   &   0.048   &  0.90   &  0.025    &  2.22   &  0.052   &  2.26   &  0.048 \\
Si/P/S   	& -0.25   &   0.001   & -0.05   &  0.000    & -0.14   &  0.003   & -0.03   &  0.002 \\
B      		& -0.25   &   0.001   & -0.06   &  0.001    & -0.22   &  0.001   & -0.23   &  0.002 \\
\hline
$\mu_{\text{tot}}$	&         &   9.20    &         &  2.42     &         &  9.15    &         &  9.59  \\
\hline \hline
\end{tabular}
\end{table}

\subsection{MAE of Fe$_5$Si$_{1-x}$P$_x$B$_2$ and Fe$_5$P$_{1-x}$S$_x$B$_2$ } 

The M\"{o}ssbauer study revealed the ''in-plane'' spin orientation for Fe$_5$SiB$_2$ below 140~K~\cite{Ericsson78}, in contrast to the ``easy-axis'' spin orientation for Fe$_5$PB$_2$ down to 80~K.
As we are interested in materials with high magnetocrystalline anisotropy, we extend theoretical inquires of Fe$_5$SiB$_2$ by the promising uniaxial case of Fe$_5$PB$_2$.
In this paper MAE is determined as the difference between total energies obtained for orthogonal spin quantization axes, MAE~=~$E_{100} - E_{001}$. Positive MAE value therefore denotes the uniaxial magnetocrystalline anisotropy along with easy axis along the [001] direction (tetragonal axis), whereas for negative MAE there is an easy-plane magnetization ($ab$-plane). 
Calculated MAE of Fe$_5$SiB$_2$ and Fe$_5$PB$_2$ are $-0.28$ and $0.35$~MJ/m$^3$ ($-140$ and $170$~$\mu$eV/f.u.), respectively.
The negative and positive signs agree well with the experimentally observed magnetic structures~\cite{Ericsson78}. 
For magnetic moment, going from Fe$_5$SiB$_2$ to Fe$_5$PB$_2$, the most noticeable change is that of the induced spin magnetic moment on $3p$-element, from $-0.25$ to $-0.14\mu_B$ (see Table~\ref{tab:mm}). The calculated total magnetic moments on Fe$_1$ and Fe$_2$ sites of Fe$_5$PB$_2$ are 1.86 and 2.27$\mu_B$/atom, respectively. They differ from experimental values $1.6\mu_B$ for Fe$_1$ and $2.2\mu_B$ for Fe$_2$, derived from the magnetic hyperfine fields~\cite{Haggstrom75}, however the mentioned experimental values are only indirect approximate estimations.

\begin{figure}[ht]
\includegraphics[trim = 80 0 0 80,clip,height=\columnwidth,angle=270]{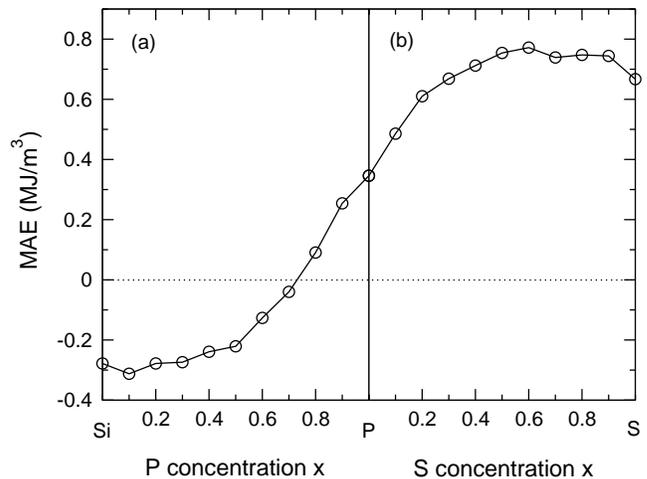}
\caption{\label{fig:MAE_vs_x} The concentration dependency of the MAE for a) Fe$_5$Si$_{1-x}$P$_x$B$_2$ and b) Fe$_5$P$_{1-x}$S$_x$B$_2$.}
\end{figure}

A priori it is not obvious, what will be the evolution of MAE as a function of concentration $x$. As an example of non-monotonic behavior we note the Fe/Co\cite{Burkert04} or (Fe/Co)$_2$B alloys\cite{Iga70}. 
Therefore we have performed VCA studies of the Fe$_5$Si$_{1-x}$P$_x$B$_2$ for the whole range of compositions, $0 \le x \le 1$. 
The resulting VCA calculations of MAE vs x are presented in Fig.~\ref{fig:MAE_vs_x}a). 
The $x$-dependence of MAE is monotonic except for the smallest values of P concentration. 
The MAE values for both stoichiometric concentrations are near extremal. 
With an increase of the P concentration the system is heading towards the uniaxial anisotropy, with a cross-over at $x \gtrsim 0.7$.
The transition from Si $3p^2$ to P $3p^3$ can be understood as an increase in the number of electrons in Fe$_5$Si$_{1-x}$P$_x$B$_2$ system.
Further increase in the number of electrons can be realized by transition from P $3p^3$ to S $3p^4$.
As shown in Fig.~\ref{fig:MAE_vs_x}b) for the hypothetical alloys with sulfur Fe$_5$P$_{1-x}$S$_x$B$_2$ significant values of MAE $\sim$0.7~MJ/m$^3$ are obtained for a broad range of sulfur concentration $0.4 \leq x \leq 1$.

\begin{figure*}[ht]
\includegraphics[trim=340 20 10 30,clip,height=\textwidth,angle=270]{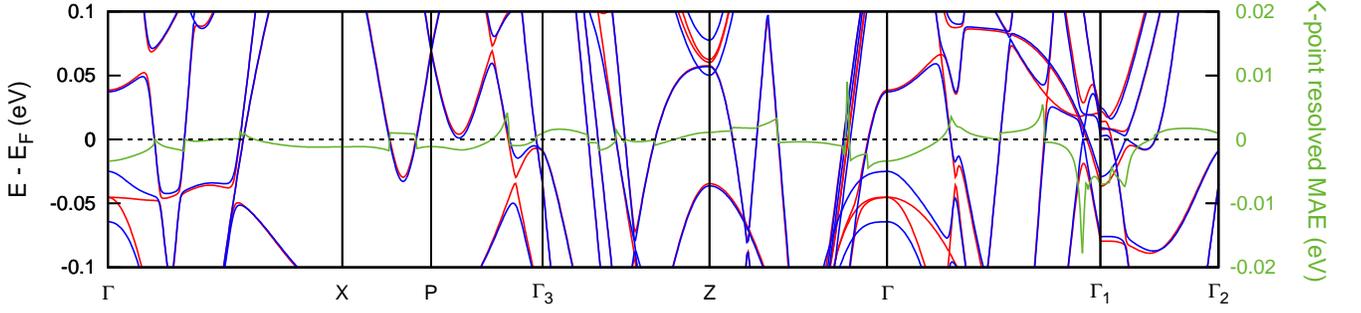}
\caption{\label{fig:fe5sib2_bands_MAE_narrow} 
Band structure of Fe$_5$SiB$_2$ calculated in fully relativistic approach for quantization axes 100 (red lines) and 001 (blue lines), together with the MAE contribution of each {\bf k}-point (green lines) as obtained by the magnetic force theorem.}
\end{figure*}


In order to get deeper insight into the origin of MAE,
the energy scale of the band structure has to be changed from tens of eV into tenths of eV, 
since the SOC constant of $3d$-metals is in the order of 0.05 eV, which is also why the spin-orbit splitting does not exceed this value, see Fig.~\ref{fig:fe5sib2_bands_MAE_narrow}.
Fully relativistic calculations result in an additional splitting of the electronic bands, due to spin-orbit coupling.
The figure presents the Fe$_5$SiB$_2$ bands in the vicinity of the Fermi level, together with the MAE contributions per $\mathbf{k}$-point as obtained by the magnetic force theorem\cite{Liechtenstein, Wu}.
Spin-orbit splitting results in different band structures for quantization axes 100 and 001 indicated by red and blue lines.

The calculated total MAE of Fe$_5$SiB$_2$ is $-140$~$\mu$eV/f.u. ($-0.28$~MJ/m$^3$).
The MAE value $\sim10^{-4}$~eV/f.u. indicates a fine compensation of the much bigger negative and positive contributions to MAE of about $10^{-3}$~eV per $\mathbf{k}$-point.
Moreover, it is easy to notice the step changes of MAE($\mathbf{k}$) whenever the band crosses the Fermi level.
Thus, when so many bands cross it, a very accurate model of the electronic band structure and fine $\mathbf{k}$-mesh are crucial to calculate MAE.

\subsection{Fe$_5$SiB$_2$ Fixed Spin Moment and Volume Dependencies}

\begin{figure}[ht]
\includegraphics[trim = 80 0 0 50,clip,height=0.9\columnwidth,angle=270]{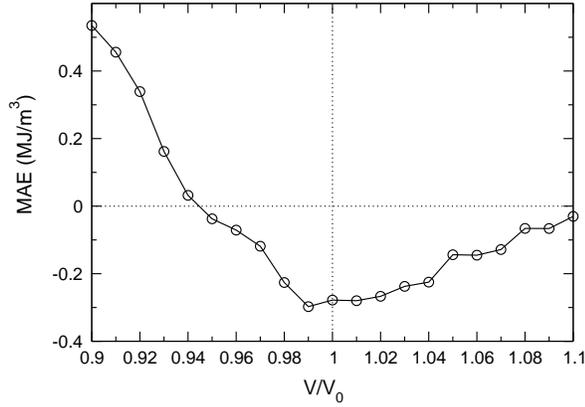}
\caption{\label{fig:MAE_vs_V} The volume dependency of the MAE for Fe$_5$SiB$_2$.}
\end{figure}

For Fe$_5$SiB$_2$ the M\"{o}ssbauer study revealed the ''in-plane'' spin orientation below 140~K and spins parallel to $c$-axis above 140~K~\cite{Ericsson78}. 
A similar transition from negative to positive magnetocrystalline anisotropy constant was observed for Fe$_2$B with a transition temperature about 520~K~\cite{Iga70}.
Since the goal of this study is to explore for potential candidates for permanent magnets, 
the experimentally revealed increase of MAE with temperature for Fe$_5$SiB$_2$ 
has motivated theoretical studies which should allow for better understanding of this phenomena.
To answer the question, whether the increase of MAE with $T$ is related to the volume expansion, we have calculated volume-dependency of MAE.
Experimentally observed increase of volume between 16 and 165~K is only about 0.01\%, but for this small volume expansion no significant change of MAE is observed, see Fig~\ref{fig:MAE_vs_V}.
The performed calculations cover a much bigger range of volume change, namely from $-10$\% to $+10$\% and,
what is striking, for a very large volume reduction the negative MAE change its sign, and for $V/V_{0} = 0.9$ reaches significant value 0.52~MJ/m$^3$. 
By a rough estimation, such volume reduction would correspond to about 20~GPa pressure, readily available in high pressure experiments~\cite{Werwinski16}.

\begin{figure}[ht]
\includegraphics[trim = 80 0 0 40,clip,height=0.9\columnwidth,angle=270]{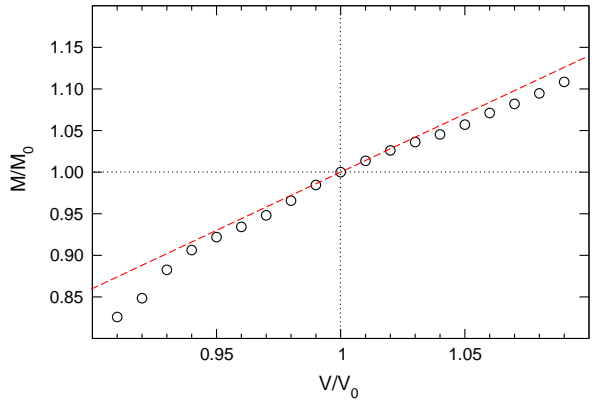}
\caption{\label{fig:M_vs_V} The volume dependency of the magnetic moment for Fe$_5$SiB$_2$ - black circles. 
For a better perception of its deviation from linearity the arbitrary (dashed) straight line is drawn. }
\end{figure}

\begin{figure}[ht]
\includegraphics[trim = 80 0 0 60,clip,height=0.9\columnwidth,angle=270]{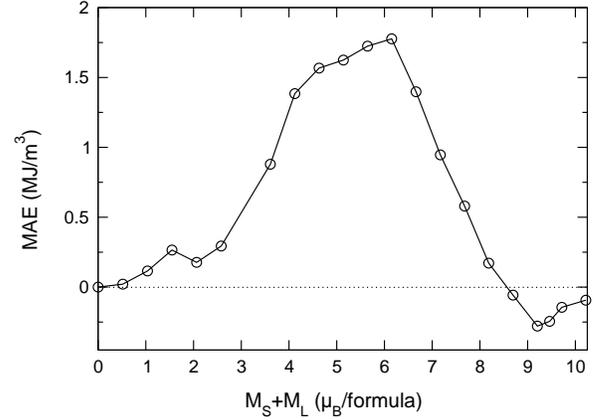}
\caption{\label{fig:MAE_vs_M_for_FSM} The total magnetic moment dependency of the MAE for Fe$_5$SiB$_2$. Based on fully relativistic fixed spin moment (FSM) calculations.}
\end{figure}

Fig.~\ref{fig:M_vs_V} shows non-linear volume dependency of the total magnetic moment for Fe$_5$SiB$_2$.
From figures \ref{fig:MAE_vs_V} and \ref{fig:M_vs_V} one could contemplate that MAE might increase when magnetic moment decreases.
To address the question, whether the MAE changes in the same manner with only magnetic moment variation and without change of volume, a set of fully relativistic fixed spin moment (FSM) calculations was performed. 
Fig.~\ref{fig:MAE_vs_M_for_FSM} presents the total magnetic moment ($M_S+M_L$) dependency of MAE. $M_S+M_L$ is evaluated for every fixed spin magnetic moment $M_S$.

FSM calculations show for Fe$_5$SiB$_2$ that MAE starts to increase with the decrease of total magnetic moment, and reaches the maximal value of MAE~=~1.78~MJ/m$^3$ for $M_S+M_L$~=~6.15$\mu_B$/f.u., while equilibrium $M_S+M_L$~=~9.20$\mu_B$/f.u.
We suggest that significant increase of MAE might occur if we would find a way to reduce the magnetic moment of Fe$_5$SiB$_2$ by 1/3.
A possible venue to achieve this could be alloying of magnetic element Fe with an element carrying a lower magnetic moment, Co.
Following this idea the VCA calculations of (Fe$_{1-x}$Co$_x$)$_5$SiB$_2$ are carried out, expecting that alloying with Co will reduce magnetic moment, and hopefully in consequence increase the MAE.
It is important to comment that reducing the magnetic moment by FSM or by alloying with another element will necessarily result in different changes to electronic structure, therefore we can't expect this analogy to hold, at least not up to large moment reduction. Nevertheless it is of interest to find out to what extent is this parallel realistic. Moreover, alloying of iron-based magnetic materials with Co has already been shown to lead to an increased MAE in several cases\cite{Burkert04,Delczeg14,Reichel,Edstrom15}.

\subsection{(Fe$_{1-x}$Co$_x$)$_5$SiB$_2$}

VCA calculations of the whole range of (Fe$_{1-x}$Co$_x$)$_5$SiB$_2$ concentrations start from optimizing the extreme structures of 
Fe$_5$SiB$_2$ and Co$_5$SiB$_2$ (see Sect.~\ref{sec:exp_comp_details} and Table~\ref{tab:crystal_data}) and interpolating crystallographic parameters for intermediate compositions.
The spin magnetic moments on Co$_1$ and Co$_2$ atoms in Co$_5$SiB$_2$ are 0.41 and 0.90$\mu_B$, respectively. They induce the small opposite spin moments on Si and B, equal to $-0.05$ and $-0.06\mu_B$, respectively. The total magnetic moment per Co$_5$SiB$_2$ formula unit is equal to 2.42$\mu_B$ (see Table~\ref{tab:mm}).

\begin{figure}[ht]
\includegraphics[trim = 80 0 0 50,clip,height=0.9\columnwidth,angle=270]{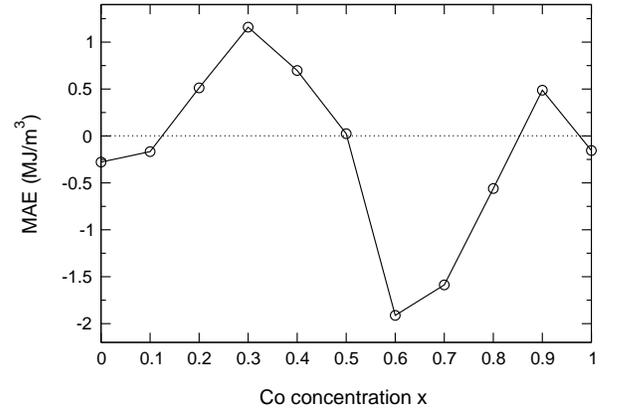}
\caption{\label{fig:Co_VCA_MAE_vs_x} The Co concentration dependency of the MAE for (Fe$_{1-x}$Co$_x$)$_5$SiB$_2$.}
\end{figure}

\begin{figure}[ht]
\includegraphics[trim = 80 0 0 50,clip,height=0.9\columnwidth,angle=270]{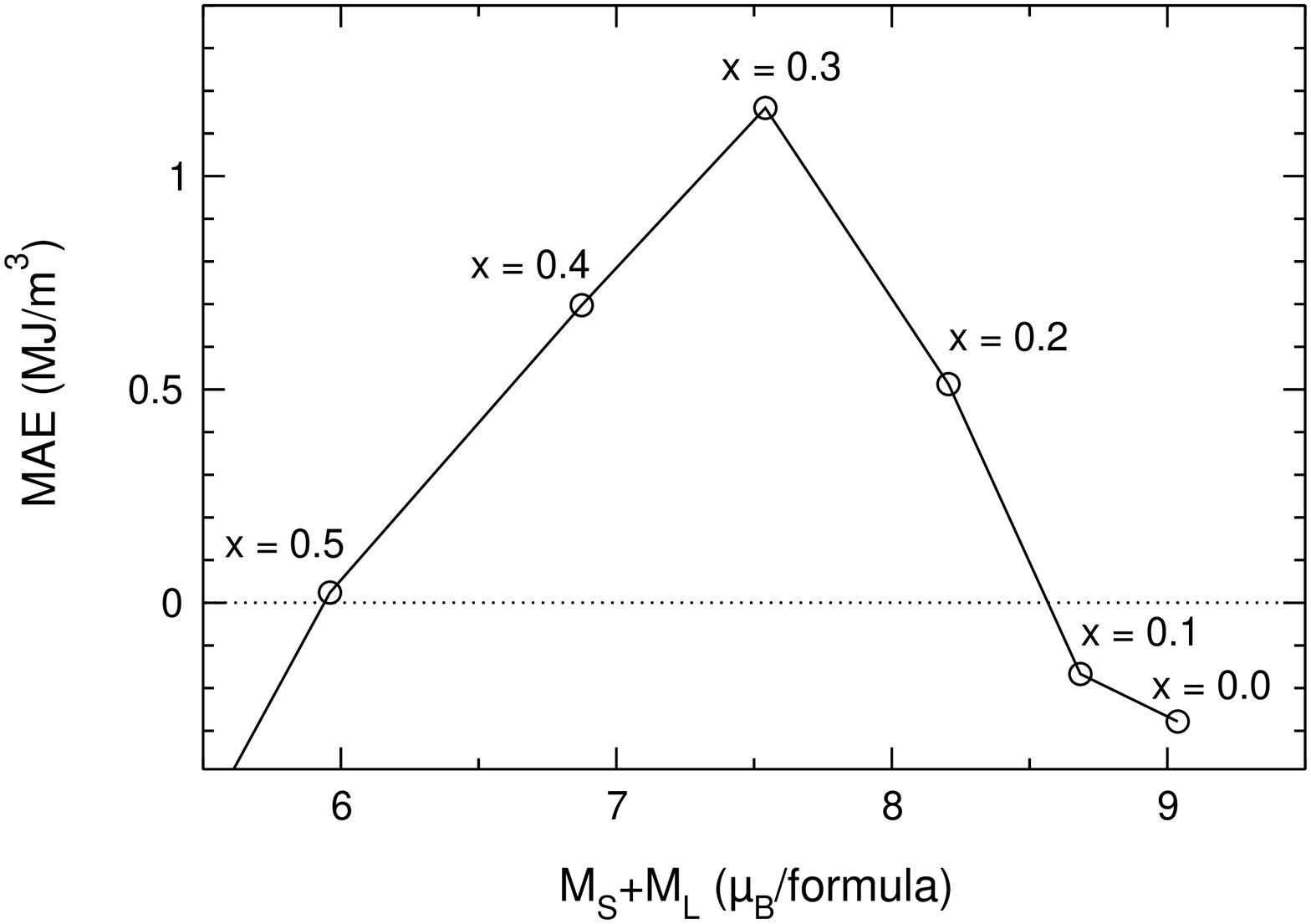}
\caption{\label{fig:Co_VCA_MAE_vs_M} The total magnetic moment dependency of the MAE for (Fe$_{1-x}$Co$_x$)$_5$SiB$_2$.}
\end{figure}

We expect that Fe/Co alloying will reduce the magnetic moment, and thus may increase MAE of Fe$_5$SiB$_2$.
The results of MAE(x) presented in Fig.~\ref{fig:Co_VCA_MAE_vs_x} confirms our predictions partially. 
The maximum of MAE~=~1.16~MJ/m$^3$ is obtained for $x = 0.3$ together with $M_S+M_L$ reduced from 9.20 to 7.75$\mu_B$/f.u., and not for 6.15$\mu_B$/f.u., as predicted for Fe$_5$SiB$_2$ from VCA.
Fig.~\ref{fig:Co_VCA_MAE_vs_M} shows explicitly the dependence of MAE on $M_S+M_L$ for (Fe$_{1-x}$Co$_x$)$_5$SiB$_2$.

\begin{figure}[ht]
\includegraphics[trim = 50 0 5 50,clip,height=0.9\columnwidth,angle=270]{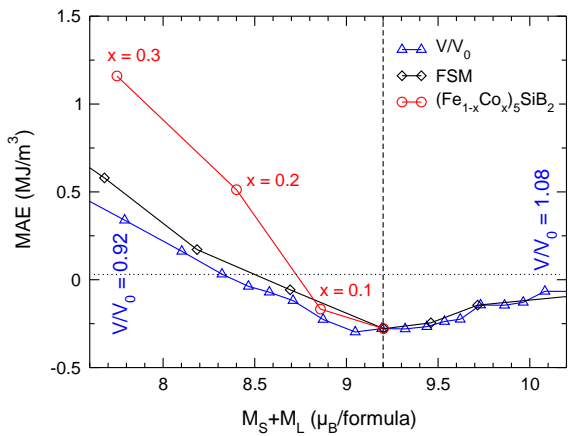}
\caption{\label{fig:Co_VCA_MAE_vs_three_M} The total magnetic moment dependencies of the MAE for Fe$_5$SiB$_2$, mediated by volume variation ($V/V_0$), fixed spin moment (FSM) and alloying with Co ((Fe$_{1-x}$Co$_x$)$_5$SiB$_2$). Dashed line denotes equilibrium state with $x=0$, $V/V_0=1$, and $M/M_0=1$.}
\end{figure}

Our studies of MAE($M_S+M_L$) are then summarized in Fig.~\ref{fig:Co_VCA_MAE_vs_three_M}, where the total magnetic moment $M_S+M_L$ was varied by changing volume, fixed spin moment (FSM) or Fe/Co concentration. 
The MAE($M_S+M_L$) variations induced by FSM or volume scaling are very similar.
The behavior of MAE($M_S+M_L$) related to alloying Fe$_5$SiB$_2$ with Co exhibit significant differences from the other two, which most probably roots in the change of the number of electrons in the system, accompanied with the Co alloying.

\begin{figure*}[ht]
\includegraphics[trim=130 20 40 40,clip,height=\textwidth,angle=270]{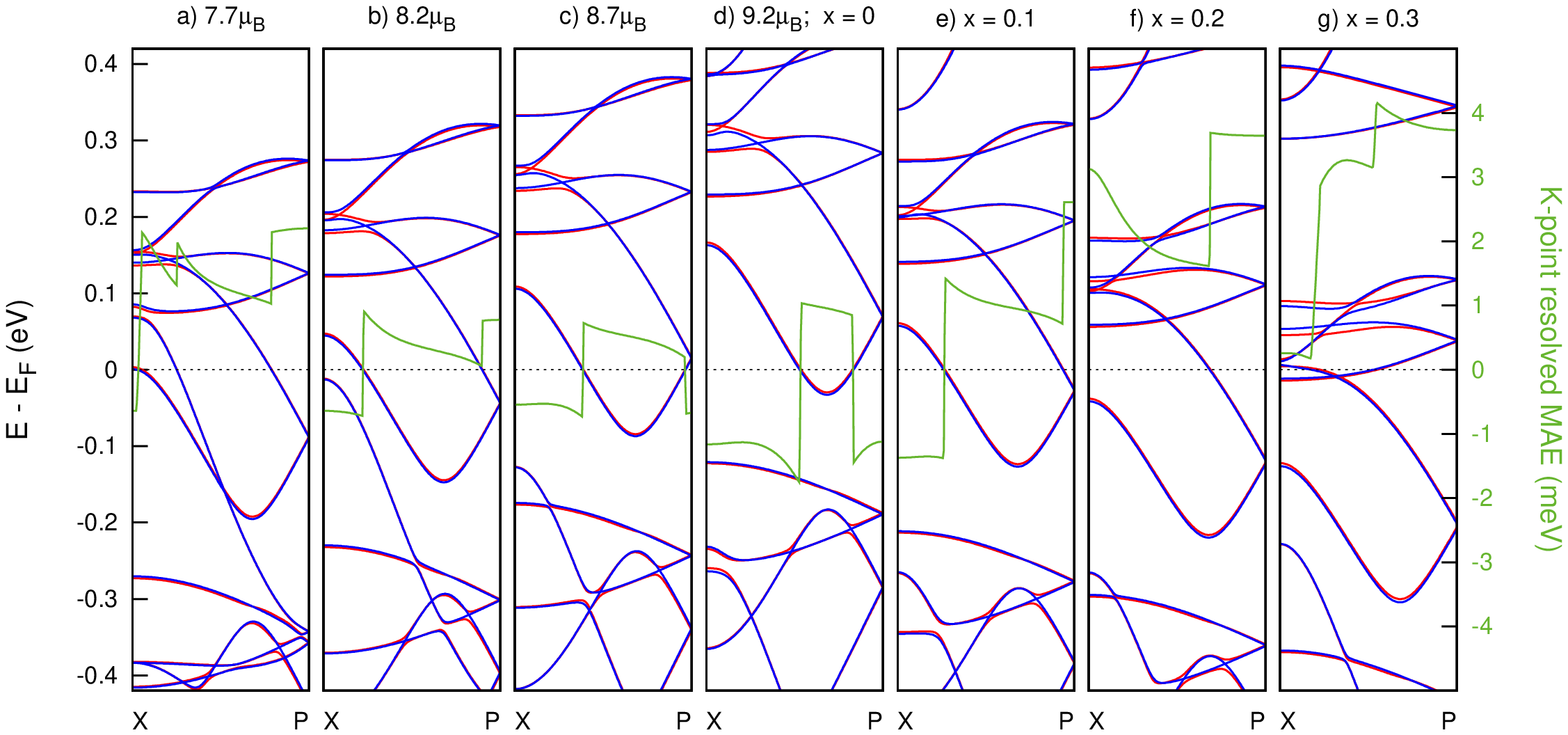}
\caption{\label{fig:feco5sib2_bandplot_comp} 
Band structures calculated in fully relativistic approach for quantization axes 100 (red lines) and 001 (blue lines), together with the MAE contributions of each {\bf k}-point (green line) as obtained by the magnetic force theorem. Panels d), c), b), and a) depict results for Fe$_5$SiB$_2$ with an increasing fixed spin moment (FSM), in representation of total moment M$_{S}$+M$_{L}$ [$\mu_B$] per formula unit. Panels d), e), f), g) present increase of Co concentration $x$ in (Fe$_{1-x}$Co$_x$)$_5$SiB$_2$ alloys.}
\end{figure*}


These differences can be examined based on the detailed band structure analysis as presented in Fig.~\ref{fig:feco5sib2_bandplot_comp}.
The energy window from $-0.5$ to $0.5$~eV allows to keep track on the variation of Fermi level with fixing spin moment FSM and doping with Co. 
Together with electronic bands the MAE contributions of each {\bf k}-point are presented as obtained by the magnetic force theorem\cite{Liechtenstein, Wu}.
The direction between high symmetry points $X$ and $P$ has been chosen because of the exceptionally simple form of bands in comparison to the other directions in Fig.~\ref{fig:fe5sib2_bands_MAE_narrow}.
For $X$-$P$ direction it is easy to notice how the band is filled with an increase of Co concentration $x$ (panels d), e), f), and g)) or a corresponding behavior induced by reduction of spin moment (panels  d), c), b), and a)).
The filling of the bands entails the change in MAE.
In the energy window from $-0.5$ to $0.5$~eV the differences in spin-orbit splitting for two quantization axes 100 and 001 might be difficult to notice. 
Nonetheless, the band structures calculated for two perpendicular quantization axes are marked in red and blue, respectively.
Differences in spin-orbit splitting can be observed on several bands crossings.
For Co concentration $x = 0$ (Fig.~\ref{fig:feco5sib2_bandplot_comp}~d)) it is particularly easy to notice that the extra positive contribution to MAE is related to the band crossing Fermi level.
For this band the 001 solution (blue line) lies slightly below the corresponding 100 band (red line), thus for the whole filled region, up to Fermi level, the positive contributions to MAE is observed.

\begin{figure}[ht]
\includegraphics[trim=0 45 0 50,clip,width=\columnwidth]{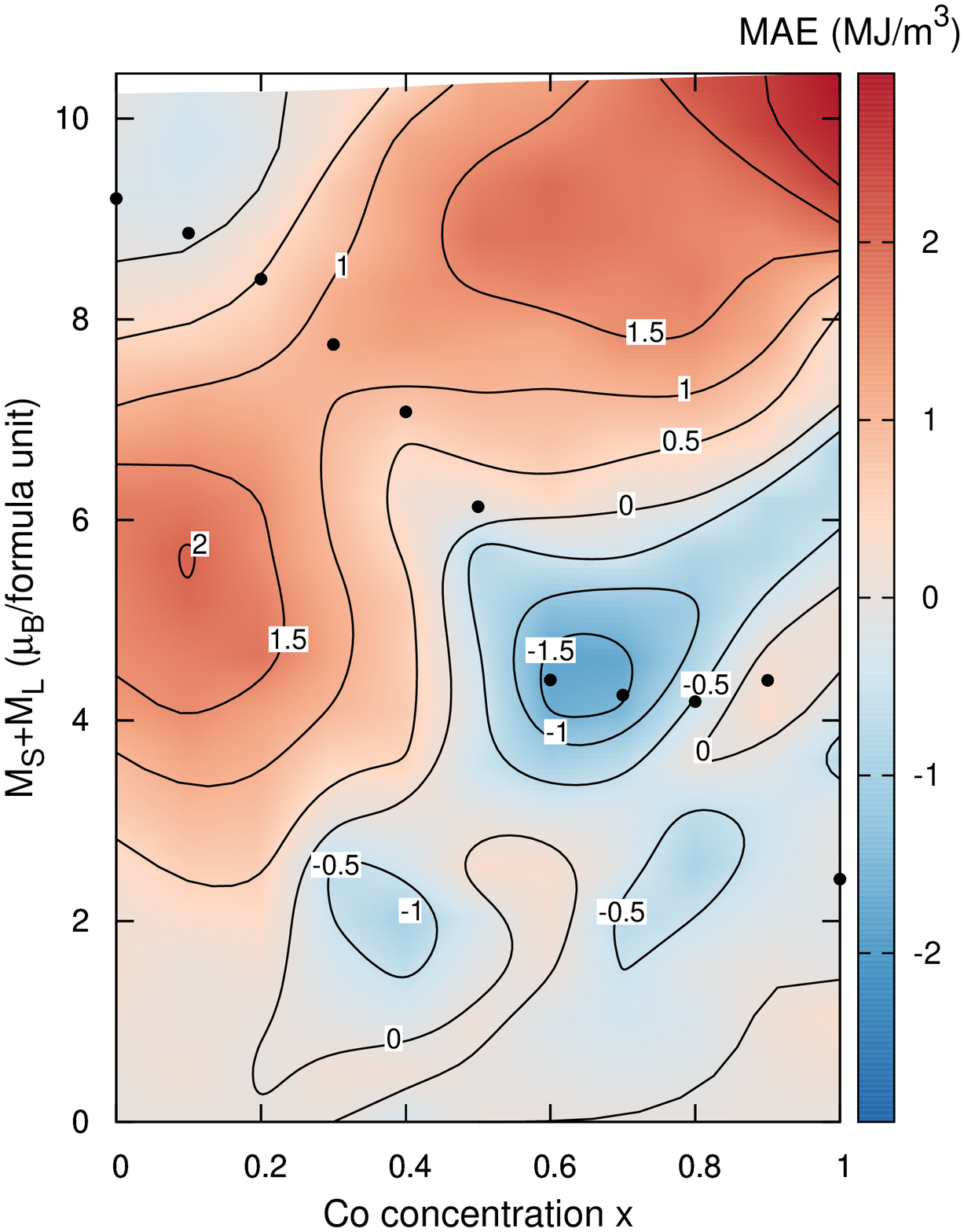}
\caption{\label{fig:feco5sib2_color_map} 
MAE as a function of $x$ and total magnetic moment (M$_{S}$+M$_{L}$) for (Fe$_{1-x}$Co$_x$)$_5$SiB$_2$.
Disorder was treated by the VCA. M$_{S}$+M$_{L}$ was stabilized with fixed spin moment (FSM) approach. Equilibrium M$_{S}$+M$_{L}$ are denoted by the black dots.
}
\end{figure}

The substantial differences in MAE(M$_{S}$+M$_{L}$) mediated by FSM or Co alloying raises the question whether this two mechanism work together.
In Fig.~\ref{fig:feco5sib2_color_map} the FSM and VCA dependencies of MAE are presented in a form of two-dimensional color map in superposition with the theoretical equilibrium total magnetic moments (M$_{S}$+M$_{L}$).
The MAE landscape reveals that by going from equilibrium magnetic moment of pure Fe$_5$SiB$_2$ towards the much lower equilibrium magnetic moment of Co$_5$SiB$_2$ the equilibrium MAE path (black dots) starts from negative sign, goes across the range of positive values and follows through a steep hollow of negative values.
This equilibrium path crossing the MAE landscape correspond to the MAE(x) function presented in Fig.~\ref{fig:Co_VCA_MAE_vs_x}.
Focusing on the uniaxial anisotropy, the highest MAE value along the equilibrium path is reached for the (Fe$_{0.7}$Co$_{0.3}$)$_5$SiB$_2$ alloy and is equal to 1.16~MJ/m$^3$.
We note that on the calculated MAE landscape, there is a region with about twice higher values of MAE for $x = 0.1$ and M$_{S}$+M$_{L}$ about 5.8$\mu_B$/f.u.
To approach this region, starting, e.g., from the (Fe$_{0.7}$Co$_{0.3}$)$_5$SiB$_2$ composition, both the magnetic moment and the $d$-electrons number (proportional to $x$) should be reduced.
The desired magnetic moment reduction, from 7.75 to around 5.8$\mu_B$/f.u., is around 25\%.
This can be obtained by alloying (Fe$_{0.7}$Co$_{0.3}$)$_5$SiB$_2$ with 25\% of a suitable non-magnetic element.
The non-magnetic alloying should at the same time decreases $x$ (which may be understood as the number of $d$ electrons beyond those of Fe) by about 0.2.
Such decrease can be obtained by alloying (Fe$_{0.7}$Co$_{0.3}$) with elements having less of $d$ electrons, i.e., the elements from columns of the periodic table preceding the column with Fe.
Among $3d$-elements which fulfill the latter condition are Cr or Mn, but these often carry rather large magnetic moments,
therefore they would not fulfill the first condition -- reduction of average moment per atom.
From $4d$- and $5d$-elements possible candidates are Mo and Tc, or W and Re, respectively.

Similar analysis of MAE landscape as presented above has been previously performed for (FeCo)$_2$B~\cite{Edstrom15} 
 and led to very similar results.
The additional calculations for (FeCo)$_2$B with W and Re confirmed that these doped systems should exhibit significant increase of MAE.
Subsequently conducted synthesis succeeded in producing (Fe$_{0.675}$Co$_{0.3}$Re$_{0.025}$)$_2$B sample for which the theoretically predicted increase of MAE has been observed~\cite{Edstrom15}.
Thus we propose that also for (Fe$_{0.7}$Co$_{0.3}$)$_5$SiB$_2$, alloying it with with W and Re might help to improve MAE. This is, however, outside the scope of the present manuscript and is left for future investigations.

The VCA yields similar results of MAE as a function of $x$ in (Fe$_{1-x}$Co$_x$)$_5$SiB$_2$ when comparing with (Fe$_{1-x}$Co$_x$)$_2$B~\cite{Edstrom15}.
Nevertheless, the VCA may quantitatively overestimate the MAE~\cite{Kota12,Turek12,Edstrom15}. 
Hence, calculations have also been attempted using the SPRKKR~\cite{Ebert11, sprkkr} method in the atomic sphere approximation (ASA) with the coherent potential approximation (CPA) to treat the alloying. However, it turns out that the SPRKKR-ASA calculations yield magnetic moments in poor agreement with the full potential calculations performed with the FPLO, as illustrated in Table~\ref{table_moments_kkr}. We suspect that the primary reason for this discrepancy is the lack of full-potential effects, since our SPRKKR calculations were done with the same exchange-correlation functional~\cite{Perdew96} and the ASA might not give an accurate description of magnetic properties in non-close packed systems~\cite{Edstrom15}. To confirm this suspicion, full potential calculations were also performed for Fe$_5$SiB$_2$ in SPRKKR and as seen in Table~\ref{table_moments_kkr}, the agreement with the FPLO is much better. 
The failure of the ASA is likely to be related to the empty spaces observed in the structure of this system~\cite{Aronsson60}, since it should expected to be a good approximation for close packed structures.
The magnetocrystalline anisotropies calculated with the CPA in SPRKKR-ASA (not shown) qualitatively disagree with 
the FPLO VCA results but as the MAE is a delicate magnetic property depending sensitively on the electronic structure, we do not consider MAE values obtained in SPRKKR-ASA to be reliable when the magnetic moments are not accurately described. Full potential MAE calculations using CPA are more challenging and beyond the scope of the current work.
\begin{table}[h]
\caption{\label{table_moments_kkr} Spin magnetic moments, in $\mu_{\text{B}}$, in Fe$_5$SiB$_2$ as calculated in the SPRKKR-ASA, SPRKKR-FP, and FPLO.
}
\begin{tabular}{ lccccc  }
\hline \hline
	   & $\text{Fe}_1$ 	& $\text{Fe}_2$ & Si 		& B 		& formula \\
\hline  
SPRKKR-ASA & 1.73 		& 2.63 		& -0.12 	& -0.15 	& 9.22 \\
SPRKKR-FP  & 1.86 		& 2.22 		& -0.15 	& -0.16 	& 9.20 \\
FPLO 	   & 1.87 		& 2.24 		& -0.25 	& -0.25 	& 8.97 \\
\hline \hline
\end{tabular}
\end{table}
We point out that VCA prediction agrees with the recent experimental results~\cite{McGuire15} of MAE increase for alloying Fe$_5$SiB$_2$ with Co.
McGuire and Parker~\cite{McGuire15} have observed for (Fe$_{0.8}$Co$_{0.2}$)$_5$SiB$_2$ (Fe$_{4}$CoSiB$_2$) the increase of anisotropy field accompanied by 20\% reduction of magnetic moment, in comparison to the pure Fe$_5$SiB$_2$.
Their measurements done on Fe$_5$SiB$_2$ are also in good agreement with our experimental results.
For Fe$_5$PB$_2$ Lamichhane et al.~\cite{Lamichhane16} report experimental values of MAE~=~0.5 MJ/m$^3$, and total magnetic moment equal to 8.6$\mu_B$/f.u.
McGuire and Parker~\cite{McGuire15} calculated in plane anisotropy with K$_1$ equal to -0.42 (-0.28) MJ/m$^3$ with total magnetic moment 9.15 (9.20) $\mu_B$/f.u. for Fe$_5$SiB$_2$, and uniaxial anisotropy with K$_1$ equal to 0.46 (0.35) MJ/m$^3$ with total magnetic moment 8.95 (9.15) $\mu_B$/f.u. for Fe$_5$PB$_2$, for comparison the results of our calculations in parenthesis. 
The differences between MAE calculated by us and by McGuire and Parker most likely come from the fact that we have used the optimized lattice parameters instead of the experimental ones.
The fact that we employed the FPLO and McGuire and Parker used the WIEN2k should not lead to substantial differences~\cite{Edstrom15}.

\section{Summary and conclusions}

We have presented an experimental study of structural and magnetic properties of Fe$_5$SiB$_2$ polycrystalline materials. This study was computationally extended across the whole range of Fe$_5$Si$_{1-x}$P$_x$B$_2$, Fe$_5$P$_{1-x}$S$_x$B$_2$, and (Fe$_{1-x}$Co$_x$)$_5$SiB$_2$ alloys, with an effort to evaluate the magnetocrystalline anisotropy energies. Theoretical study of volume variation and fully relativistic fixed spin moment calculations of stoichiometric Fe$_5$SiB$_2$ evidences a strong inverse dependency between MAE and total magnetic moment, leading to a maximal value of MAE~=~1.78~MJ/m$^3$ for $M_S+M_L$~=~6.15$\mu_B$/f.u., while equilibrium magnetic moment is 9.20$\mu_B$/f.u. 
Alloying of Fe$_5$SiB$_2$ with Co is suggested to reproduce reduction of magnetic moment. 
The whole range of (Fe$_{1-x}$Co$_x$)$_5$SiB$_2$ concentrations is calculated and evidence of a MAE maximum is obtained for Co concentration $x=0.3$, with the value of MAE~=~1.16~MJ/m$^3$ and with the magnetic moment 7.75$\mu_B$/f.u. 
Thus, (Fe$_{0.7}$Co$_{0.3}$)$_5$SiB$_2$ appears to be a promising candidate for a rare-earth free permanent magnet.
Also further increase of MAE of (Fe$_{0.7}$Co$_{0.3}$)$_5$SiB$_2$ is expected by doping with some of $4d$- or $5d$-elements.
For Fe$_5$Si$_{1-x}$P$_x$B$_2$ a monotonic trend in MAE has been obtained, suggesting that at low temperatures and below $x \approx 0.8$ the alloys should have in-plane magnetization, while above $x \approx 0.8$ the magnetization should point along the $c$-axis. The overall MAE values are rather low, suggesting that Fe$_5$Si$_{1-x}$P$_x$B$_2$ class of materials can only form soft magnets.
On the other hand the hypothetical Fe$_5$P$_{1-x}$S$_x$B$_2$ compositions with sulfur exhibit increased MAE with the highest value of 0.77~MJ/m$^3$ for Fe$_5$P$_{0.4}$S$_{0.6}$B$_2$ and  MAE~=~0.67~MJ/m$^3$ for Fe$_5$SB$_2$.

\section{Acknowledgements}

We gratefully acknowledge the financial support of G\"{o}ran Gustafsson's Foundation, Swedish Research Council and an EU-FP7 project REFREEPERMAG.


\begin{thebibliography}{}

\bibitem{Burkert04} T. Burkert, L. Nordstr\"om, O. Eriksson, and O. Heinonen, {\em Phys. Rev. Lett.}~\textbf{93} (2004) 027203.

\bibitem{Andersson06} G. Andersson, T. Burkert, P. Warnicke, M. Bj\"orck, B. Sanyal, C. Chacon, C. Zlotea, L. Nordstr\"om, P. Nordblad, and O. Eriksson, {\em Phys. Rev. Lett.}~\textbf{96} (2006) 037205.

\bibitem{Delczeg14} E. K. Delczeg-Czirjak, A. Edstr\"{o}m, M. Werwi\'{n}ski, J. Rusz, N. V. Skorodumova, L. Vitos, and O. Eriksson, {\em Phys. Rev. B.}~\textbf{89} (2014) 144403.

\bibitem{Reichel} L. Reichel, G. Giannopoulos, S. Kauffmann-Weiss, M. Hoffmann, D. Pohl, A. Edström, S. Oswald, D. Niarchos, J. Rusz, L. Schultz, and S. Fähler, 
{\em J. Appl. Phys.}~\textbf{116} (2014) 213901.

\bibitem{Reichel15} L. Reichel, L. Schultz, D. Pohl, S. Oswald, S. Fähler, M. Werwiński, A. Edström, E. K. Delczeg-Czirjak, and J. Rusz, {\em J. Phys. Cond. Matt.}~\textbf{27} (2015) 476002.

\bibitem{Miura13} Y. Miura, S. Ozaki, Y. Kuwahara, M. Tsujikawa, K. Abe and M. Shirai, {\em J. Phys. Cond. Matt.}~\textbf{25} (2013) 106005.

\bibitem{Edstrom14} A. Edstr\"{o}m, J. Chico, A. Jakobsson, A. Bergman, and J. Rusz, {\em Phys. Rev. B}~\textbf{90} (2014) 014402.

\bibitem{Kim72} T. K. Kim, M. Takahashi, {\em Appl. Phys. Lett.}~\textbf{20} (1972) 492.

\bibitem{Iga70} A. Iga, {\em Japan. J. Appl. Phys.}~\textbf{9} (1970) 415.

\bibitem{Belashchenko15} K.~D.~Belashchenko, L.~Ke, M.~D\"{a}ne, L.~X.~Benedict, T.~N.~Lamichhane, V.~Taufour, A.~Jesche, S.~L.~Bu\'{d}ko, P.~C.~Canfield, V.~P.~Antropov, {\em Appl. Phys. Lett.}~\textbf{106}, 062408 (2015).

\bibitem{Edstrom15} A. Edström, M. Werwiński, D. Iuşan, J. Rusz, O. Eriksson, K. P. Skokov, I. A. Radulov, S. Ener, M. D. Kuz’min, J. Hong, M. Fries, D. Y. Karpenkov, O. Gutfleisch, P. Toson, and J. Fidler, {\em Phys. Rev. B}~\textbf{92} (2015) 174413.

\bibitem{Fujii77} H. Fujii, T. Hokabe, T. Kamigaichi and T. Okamoto, {\em J. Phys. Sosc. Jpn.}~\textbf{43} (1977) 41.

\bibitem{Delczeg10} E. K. Delczeg-Czirjak, L. Delczeg, M. P. J. Punkkinen, B. Johansson, O. Eriksson, and L. Vitos, {\em Phys. Rev. B}~\textbf{82} (2010) 085103.

\bibitem{Niarchos15} D. Niarchos, G. Giannopoulos, M. Gjoka, C. Sarafidis, V. Psycharis, J. Rusz, A. Edström, O. Eriksson, P. Toson, J. Fidler, E. Anagnostopoulou, U. Sanyal, F. Ott, L.-M. Lacroix, G. Viau, C. Bran, M. Vazquez, L. Reichel, L. Schultz, and S. Fähler, {\em JOM}~\textbf{67} (2015) 1318.

\bibitem{Aronsson60} B. Aronsson, I. Engstr\"{o}m, {\em Acta Chemica Scandinavica}~\textbf{14} (1960) 1403-1413.

\bibitem{Rundqvist62} S. Rundqvist, {\em Acta Chemica Scandinavica}~\textbf{16} (1962) 1-19.

\bibitem{Reis07} D.A.P. Reis, C.A. Nunes, A. Capri Neto, {\em Revista Brasileira de Aplica\c{c}\~{o}es de V\'{a}cuo}~\textbf{26} (2007) 79-82.

\bibitem{Brauner09} A. Brauner, C.A. Nunes, A.D. Bortolozo, G. Rodrigues, A.J.S. Machado, {\em Solid State Commun.}~\textbf{149} (2009) 467. 

\bibitem{Fukuma11} M. Fukuma, K. Kawashima, M. Maruyama, J. Akimitsu, {\em J. Phys. Soc. Jpn.}~\textbf{80} (2011) 024702.

\bibitem{Baurecht71} H.E. Baurecht, H. Boller, H. Nowotny, {\em Monatshefte f\"{u}r Chemie / Chemical Monthly}~\textbf{102} (1971) 373-384.

\bibitem{Ilnitskaya85} O. N. Ilnitskaya, Yu. B. Kuzma, {\em Soviet Powder Metallurgy and Metal Ceramics}~\textbf{24} (1985) 226-228.

\bibitem{Ericsson78} T. Ericsson, L. H\"{a}ggstr\"{o}m and R. W\"{a}ppling, {\em Physica Scripta}~\textbf{17} (1978) 83-86.

\bibitem{Haggstrom75} L. H\"{a}ggstr\"{o}m, R. W\"{a}ppling, T. Ericsson, Y. Andersson, S. Rundqvist, {\em J. Sol. St. Chem.}~\textbf{13} (1975) 84-91.

\bibitem{McGuire15} M.A. McGuire and D.S. Parker, {\em J. Appl. Phys.}~\textbf{118} (2015) 163903.

\bibitem{Lamichhane16} T.N. Lamichhane, V. Taufour, S. Thimmaiah, D. S. Parker, S. L. Bud’ko, and P. C. Canfield, {\em J.~Mag.~Magn.~Mat.}~\textbf{401} (2016) 525.

\bibitem{Rietveld69} H.M. Rietveld, {\em J. Appl. Cryst.}~\textbf{2} (1969) 65.

\bibitem{Rodriguez93} J. Rodriguez-Carvajal, {\em Physica B}~\textbf{192} (1993) 55-69.

\bibitem{Holland97} T.J.B. Holland, S.A.T. Redfern, {\em Mineral. Mag.}~\textbf{61} (1997) 65-77.

\bibitem{Koepernik99} K.~Koepernik and H.~Eschrig, {\em Phys.~Rev.~B}~\textbf{59} (1999) 1743.

\bibitem{Perdew96} J.P.~Perdew, K.~Burke and M.~Ernzerhof, {\em Phys.~Rev.~Lett.} \textbf{77} (1996) 3865.

\bibitem{Soven70} P. Soven, {\em Phys. Rev. B} \textbf{2} (1970) 4715.

\bibitem{Turek12} I. Turek, J. Kudrnovsk\'{y}, and K. Carva, {\em Phys. Rev. B}~\textbf{86} (2012) 174430.

\bibitem{Kota12} Y. Kota and A. Sakuma, {\em Appl. Phys. Express}~\textbf{5} (2012) 113002.

\bibitem{Momma08} K. Momma and F. Izumi, {\em J. Appl. Crystallogr.} \textbf{41} (2008) 653.

\bibitem{Chen95} C.T. Chen, Y.U. Idzerda, H.-J. Lin, N.V. Smith, G. Meigs, E. Chaban, G.H. Ho, E. Pellegrin, and F. Sette, {\em Phys.~Rev.~Lett.} \textbf{75} (1995) 152.

\bibitem{Cedervall16} J. Cedervall, S. Kontos, T. C. Hansen, O. Balmes, F. J. Martinez-Casado, Z. Matej, P. Beran, P. Svedlindh, K. Gunnarsson, and M. Sahlberg, {\em Phys. Rev. B} \textbf{235} (2016) 113.

\bibitem{Liechtenstein} A.I.~Liechtenstein, M.I.~Katsnelson, V.P.~Antropov, and V.A.~Gubanov, {\em J.~Mag.~Magn.~Mat.}~\textbf {67} (1987) 65-74.

\bibitem{Wu} R. Wu, and A.J. Freeman, {\em J.~Mag.~Magn.~Mat.}~\textbf{200} (1999) 498-514.

\bibitem{Werwinski16} M. Werwiński, J. Kaczkowski, P. Leśniak, W. L. Malinowski, A. Szajek, A. Szczeszak, and S. Lis, {\em Comp. Mat. Sci.} \textbf{117} (2016) 98.

\bibitem{Ebert11} H. Ebert, D. K\"{o}dderitzsch, and J. Min\'{a}r, {\em Rep.~Prog.~Phys.}~\textbf{4} (2011) 096501.

\bibitem{sprkkr} H. Ebert, {The munich spr-kkr package, version 7.2}, http://ebert.cup.uni-muenchen.de/SPRKKR


\end{thebibliography}
\end{document}